\documentclass{JINST}

 \usepackage{subfigure}
 \usepackage{graphics}
 \usepackage{graphicx}
\usepackage{lscape} %for tables ...
\usepackage{rotating}
\usepackage{amssymb}
\usepackage{amsmath}
\title{The Waveform Digitiser of the Double Chooz Experiment: Performance and Quantisation Effects on PhotoMultiplier Tube Signals}

\author{
Y.~Abe$^d$,
T.~Akiri$^c$, 
A.~Cabrera$^a$,
B.~Courty$^a$,
J.~V.~Dawson$^a$\thanks{Corresponding author.}~,
L.~F.~G.~Gonzalez$^{a,b}$,
A.~Hourlier$^a$,
M.~Ishitsuka$^d$,
H.~de~Kerret$^a$, 
D.~Kryn$^a$,
P.~Novella$^a$,
M.~Obolensky$^a$,
S.~Perasso$^a$,
A.~Remoto$^{a,e}$,
R.~Roncin$^a$\\
\llap{$^a$}Laboratoire Astroparticule et Cosmologie, 
10 rue Alice Domon et L\'eonie Duquet, 75205 Paris, France\\
\llap{$^b$}Universidade Estadual de Campinas-UNICAMP,
Campinas, SP, Brazil \\
\llap{$^c$}Duke University, Department of Physics,
 Durham, North Carolina, U.S.A\\
\llap{$^d$}Department of Physics, Tokyo Institute of Technology,
 Tokyo, 152-8551, Japan\\
\llap{$^e$}Laboratoire d'Annecy-le-Vieux de physique des particules, 9 Chemin de Bellevue, 74941 Annecy-le-Vieux, France\\

E-mail: \email{jaime.dawson@gmail.com}}

\abstract{

We present the waveform digitiser used in the Double Chooz experiment.
We describe the hardware and the custom-built firmware specifically developed
for the experiment.  The performance of the device is tested with regards to
digitising low light level signals from photomultiplier tubes and measuring
pulse charge.  This highlights the role of quantisation effects and leads to
some general recommendations on the design and use of waveform digitisers.
}


\begin{document}

% main text
\section{Introduction}
\label{intro}
The growing usage of Waveform Digitisers in Experimental Physics applications
has been made possible by the progress of Flash ADC chip industry. It is driven
by two motivations:
first, instead of recording separately some of the properties of the signal like
timing, amplitude and charge, one can now use one single digitiser per channel to
record the whole signal itself and derive its properties by software or in firmware;
second, the event represented by the signal can easily be stored
temporarily before the decision is taken to keep or reject it.  In addition, recording signal profiles allows to exploit Pulse Shape Discrimination techniques leading to a better understanding of background events, and gives the ability to discriminate better between physics and spurious signals.

The Double Chooz experiment measures the third neutrino mixing angle
$\theta_{13}$ using anti-neutrinos emitted from a nuclear power plant \cite{DC1,DC2}.  The detector uses liquid scintillators
\cite{scintillator} observed by 390 ten-inch low background PhotoMultiplier Tubes (PMT)
(Hamamatsu R7081 \cite{hamamatsu, pmt1,pmt2}) for
the neutrino target, and 78 eight-inch PMTs (Hamamatsu R1408 \cite{drexelnote})
for the muon veto. The PMTs are operated in high gain mode since the number of photons impacting on each individual PMT is low.  For an event, the charge contained in each PMT waveform is measured and the total charge observed by all PMTs is used to determine the energy deposited.  The energy range of the neutrino signal begins below 0.7 MeV for the positron interaction and extends up to 10 MeV to well contain the gamma rays resulting from neutron capture by Gadolinium.  In practice, data is recorded from 300 keV on the low energy side.  On the high energy side, the detector response is linear to approximate 50 MeV which allows sampling of important backgrounds.  In this energy range,  each PMT signal contains from zero to $\sim$50 photoelectrons.   Cosmic ray muons crossing the detector deposit even higher energies, saturating the waveform digitiser channels described in this article resulting in an overall non-linear energy response which extends up to $\sim$ 600 MeV. The recording of the PMT waveforms is therefore an important factor in the overall energy response of the detector.

In this article, we give a general description of the electronics chain involved in the digitisation of the
PMT signals (Section \ref{DC_DAQ}); then discuss the Flash ADC hardware and
describe the main features of the firmware written specifically for the experiment (Section
\ref{V1721}). Before being installed at the experimental site, a series of tests were made
on each waveform digitiser card to ensure their good working condition \cite{Akiri2010}. The results of one such test, testing the linearity of all channels, are reported in Section \ref{dnl}. The
response to PMT signals was explored with tests of a single channel mock-up of
the full Double Chooz electronics chain.  Signals cannot be perfectly recorded; there are
undesirable effects such as high frequency noise (Section \ref{noise}), ADC non-linearity,
and digitisation effects which can produce subtle biases on the measurement of pulse charge.
In Section \ref{quantisation}, we discuss the sources of bias on the measurement of pulse charge.
We show that these effects can be well reproduced by simulation, and also measured in-situ with
a suitable light calibration system. We also discuss the working conditions necessary to minimise
their effects. Finally Section \ref{simulation} shows results from a detailed simulation of a
waveform digitiser.

\section{Double Chooz Electronics}
\label{DC_DAQ}

\begin{figure}
\centering{
\includegraphics[width=12cm]{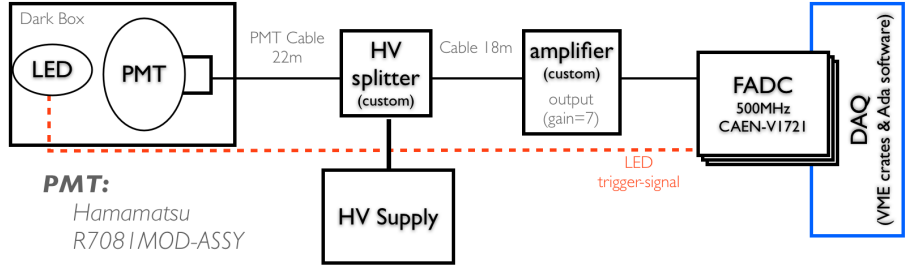}
\caption{Electronics for the single channel test.}
\label{fig:electronics}
} % caption for the whole figure
\end{figure}

Figure \ref{fig:electronics} is a diagram of the single channel electronics.
The PMTs are operated at a gain of $10^7$, the fast signal is
decoupled from the PMT HV cable by a custom-built HV splitter.  The signal
from the PMT is amplified by the Double Chooz amplifier with a gain of $\sim$7 such that after amplification the
mean single photoelectron amplitude is $\sim$ 35 mV.  The analog RMS noise
level was measured as $\sim$1.2 mV at the output of the amplifier. All PMT
signals are sampled synchronously every 2 ns by the waveform digitisers. 
The Data Acquisition system runs via VME bus, with each crate controlled by a
single-board computer (MVME3100 PowerPC \cite{mvme3100}) running Debian Linux with Data Acquisition software written in Ada 2005.

The first reason to use Waveform Digitisers for the Double Chooz experiment was
the desire to keep a full control on the determination of the signal properties.
This means to be able to control, from the recorded data, the pulse position
within the integration window, and check that the time measurement is not done
on a spurious signal. Waveform digitisers also allow to obtain simultaneous
digitisation and read out without dead time.

\section{The Double Chooz waveform digitiser}
\label{V1721}

  \begin{table}[hbp]
 \begin{center}

    \caption{Main characteristics of the Double Chooz waveform digitiser}
    \begin{tabular}{|l|c|}
      \hline 
      Time resolution & 2 ns\tabularnewline
      \hline 
      Time precision & $<$5 ps\tabularnewline
      \hline 
      Amplitude resolution (LSB) & 78 $\mu$A / 3.9 mV \tabularnewline
      \hline 
      Dynamic range & 8-bit / 20mA / 1V \tabularnewline
      \hline 
      Differential Non-Linearity & Typical maximum of |0.16| LSB \tabularnewline
                                 & with limits $\pm$ 0.6 LSB \tabularnewline
      \hline 
      Integral Non-Linearity & Typical maximum |0.3| LSB \tabularnewline
                             & with limits     $\pm$ 0.9 LSB\tabularnewline
       \hline   
      SRAM Memory Buffer Size   & 2 MiByte/channel \tabularnewline
       \hline
    \end{tabular}
  \label{WFDdatasheet}
 \end{center}
 \end{table}

The waveform digitiser for the Double Chooz experiment was developed
in partnership with CAEN, and is available as model Vx1721 (VME64x)
or V1721\cite{CAENV1721} (legacy VME). It is based on an 8-Channel 8-bit,
500 MS/s Flash ADC (FADC).  For interest, please note that a 10-bit 1Gs/s
version is now available \cite{Acciarri2012}.

The model used by Double Chooz
(see Table \ref{WFDdatasheet} for main characteristics) has single ended
signal inputs each with a dynamic range of approximately 1 Volt. A DC-offset
can be applied individually to each channel through the use of an onboard
16-bit DAC. For the negatively-going PMT signals of Double Chooz this
offset is set such that ground is found around ADC code 210, in this way
allowing some room for the digitisation of positive over-shooting signals
 whilst maintaining a good dynamic range for the
negatively-going PMT signals.

Double Chooz has developed a firmware for this board, which is different
to the commercial version developed by CAEN, and is tailored for the experiment.
The description below of the board functionalities is essentially the
description of this firmware. Each channel has a SRAM memory buffer (of total
size 2 $Mi\mbox{Byte}$\footnote{$Mi \equiv \times 2^{20} \sim 1.05\times 10^6$ })
which is divided into 1024 circular memory buffers or {\em pages}. A sophisticated
control logic divides memory access periodically in two different time slots,
one for write and one for read, only delaying the pending read or write accesses.

The data stream is continuously written into one page until the arrival
of a trigger signal.  When the trigger occurs, the page is frozen and
the acquisition continues without dead time by writing into the following
empty page. The pages are unfrozen by the means of VME commands. All
pages are permanently accessible for reading through the VMEbus. The memory
can be seen as a transparent FIFO: events enter the FIFO instantly on trigger
and are removed instantly by a VME command. In the mean time, they can be
inspected and read randomly (transparency of the FIFO). If the read-out
program was not able to unfreeze the pages fast enough, so that the FIFO
becomes full, dead-time would occur. This condition is detectable by software
which would then raise a warning or an exception. %

The front-panel clock input is used for the synchronisation of
multiple cards with the Data Acquisition clock.  Multiple cards can
therefore be triggered simultaneously by NIM signals. 
Also on the front-panel are 16 programmable LVDS Input/Outputs.

The VME capabilities of the board which are implemented in the firmware
include the fastest VME64x transfer protocol, 2eSST, at a maximum speed of
320 $Mi\mbox{Byte}/s$.
Geographical Addressing, available in VME64x, greatly facilitates the
management of a large number of boards, allowing to automate board
detection and address assignments.

\subsection{Synchronisation}
The synchronisation happens in two steps:
\begin{enumerate}
\item{\bf Card synchronisation:}
The Trigger System distributes a 62.5 MHz (16 ns period) clock and a
synchronous trigger signal. The clock is distributed as LVDS and the
trigger as NIM.

\item{\bf Channel synchronisation:}
An onboard PLL\footnote{Phase-Locked Loop} produces a 500 MHz
(2 ns period) clock synchronous with the external 62.5 MHz. The trigger
signal is distributed to each of four memory-management logical units
(one per two channels), which all switch
pages at the same time, on  edges of the 62.5 MHz clock.
\end{enumerate}

The 62.5 MHz clock gives the actual pace of the board logic. Every
16 ns, eight FADC samples (eight measurements of the input current),
done at 2 ns intervals, are stored in the SRAM memory. The maximum waveform length available is 4 $\mu$s
(2 $kiB$ per channel).  The trigger point of each board is
completely flexible, similar to the horizontal time-offset on an oscilloscope, a configurable delay can be added to move the
trigger point from the end of the waveform to the start, in steps of 16 ns.

\subsection{Trigger and Event Metadata}
The Double Chooz Trigger generates information about the triggering
conditions and calculates the event number. Synchronous to the trigger,
this trigger data is passed to the Waveform Digitiser boards
through the 16-bit LVDS input connector present on the front-panel.
This information, together with the number of clock ticks since
the previous trigger and an internal trigger counter, are stored by
the Waveform Digitisers and one set is kept for each page. This
constitutes the event {\em metadata}; it can be read from the VME,
and used to categorise the event types and differentiate data handling.

\subsection{Real-Time Signal Counting}
As the FADC is continuously digitising it can also be setup to monitor
the signal rate on each channel. This is achieved by setting (by software)
a threshold on each channel such that when it is crossed by a signal,
a positive square pulse of firmware-determined width and amplitude is
generated.  All such signals from each channel are combined and the
result is sent to the LEMO output on the front panel
through a 16-bit DAC, with an amplitude which is set in the
firmware by selecting which 4 bits out of the 16 are used.
In this way the number of channels simultaneously firing is coded into the amplitude of an analog signal.  This information per card could be used to form the
system-wide Trigger condition. This is the basis of a future enhancement
to the Double Chooz Trigger system.
%%%%%%%%%%%%%%%%%%%%%%%%%%%%%%%%%%%%%%%%%%%%%%%%%%%%%%
%
\section{FADC Testing}
\label{tests}
The main objective of the Double Chooz waveform digitiser is to record
scintillator pulses and reconstruct the contained charge. In this section
we describe the behaviour of the waveform digitiser, with particular
emphasis on the sources of bias on the charge measurement. 

FADCs convert analog waveforms to digital form, by using a linear voltage
ladder with comparators at each rung to compare input voltages to successive
reference voltages.  The output of the comparators are fed into a digital
encoder which outputs binary values.  The terms 'ADC code' or 'ADC count' will be used hereafter to denote these output values 0 to 255 for the 8-bit FADC.  The analog voltage to digital code
(or ADC) transfer function is not perfect since it relies on real electronic
components. An assessment of the linearity of all 67 FADC cards
was made and is reported in Section \ref{dnl}. This gives a general limit
on the achievable linearity performance for a single channel.

We note the presence of a low amplitude high frequency noise shown in Section
\ref{noise}, which could also affect the charge measurement.

The choice of dynamic range, signal amplitude and noise level, has an impact
on the overall charge linearity achievable. This is an important issue for
the 8-bit FADC.  This is discussed in Section \ref{quantisation} and
illustrated with tests with a single channel of the Double Chooz electronics,
including PMT pulsed with an LED.  Measurements of the gain with different
light levels, gains and noise levels, were made using the well known
photostatistics technique described in\cite{Lombard&Martin}.

\subsection{FADC Linearity}
\label{dnl}

There are two values important in the definition of
the linearity of an ADC; the first, is the Differential Non-Linearity
(DNL), and the second is the Integral Non-Linearity (INL). These
parameters determine the voltage-to-ADC transfer function, how an analog
voltage input is converted to a digital code (or ADC value).

The DNL is the measure of the deviation from the ideal step size of 1 ADC count (or Least Significant Bit).  The DNL for each ADC count can
be positive (so the step size is larger than ideal) or negative (so the step
is smaller than ideal). The INL is the cumulative effect of the $DNL$. It is
the difference between the FADC and an ideal voltage-to-ADC transfer function. For a good discussion of FADC linearity see for example \cite{DNLINL, DNLINL2}.

Measurements of the DNL for each code of 536 FADC channels were made using a simple
histogram method. A 1V 12 bit DAC, provided by one FADC card, was used as a controllable source of
DC voltage giving 4096 incremental steps of 244 $\mu$V, into the input of each FADC channel. For each voltage
input, a waveform of 1024 samples was recorded. The input voltages span a range
between 2mV and 1.02V, allowing measurements of a large fraction of each FADC
channel range (from ADC code 20 to 245). The short step voltage (244 $\mu$V)
resulted in the repeated sampling of all ADC codes in the measured range.  The
4096 waveforms recorded were used to make a histogram of the sampled ADC codes.
A perfect FADC would show a uniform sampling of all ADC codes, distortions to
the histogram indicates a variation in the code width.  In this way the DNL per
code was estimated.  As the INL is the cumulation of each DNL per code, the INL
per code was calculated from this data. 

\begin{figure}
%\centering{

\subfigure{
  \includegraphics[width=7cm]{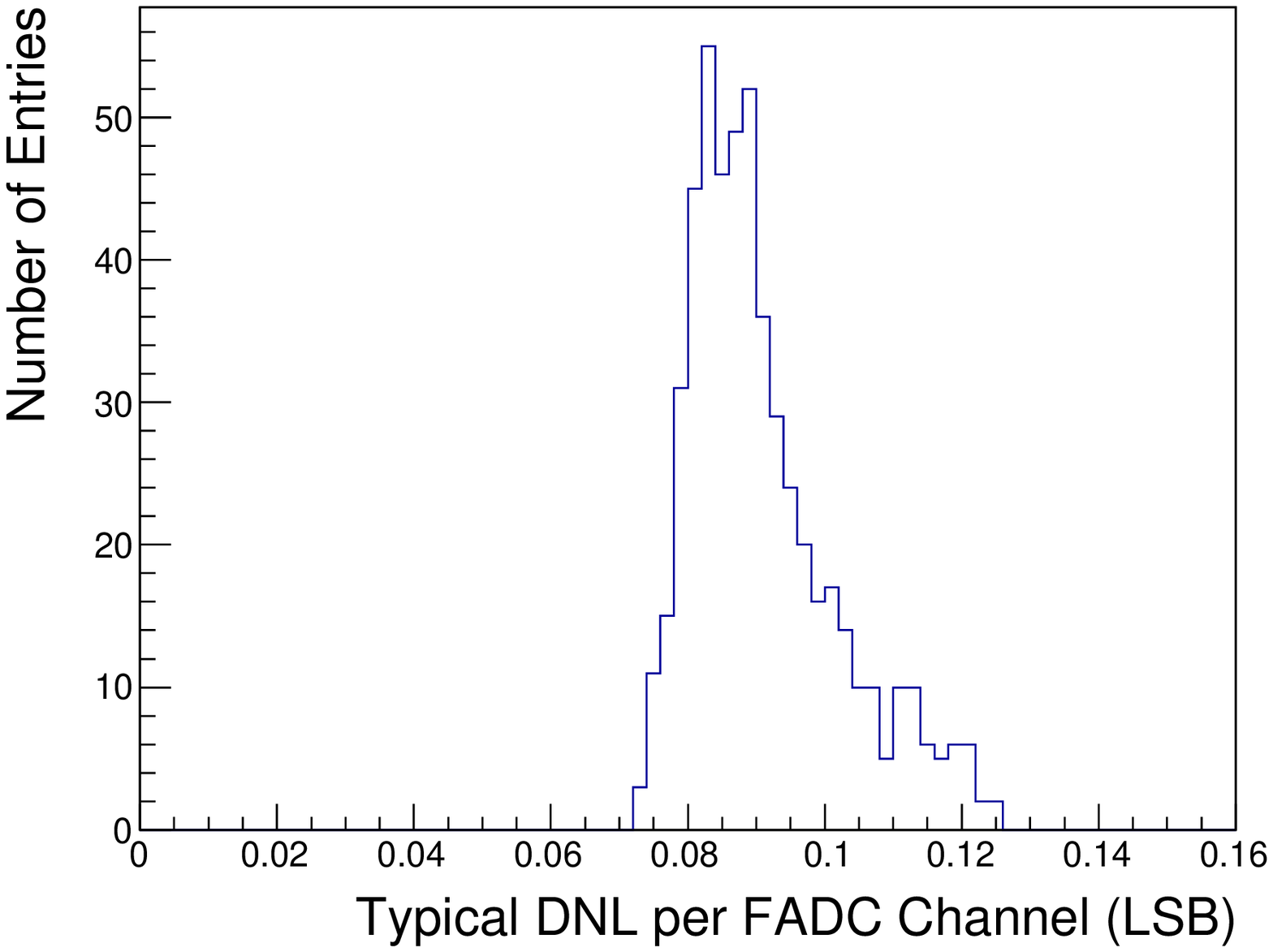}
 
}
\subfigure{
  \includegraphics[width=7cm]{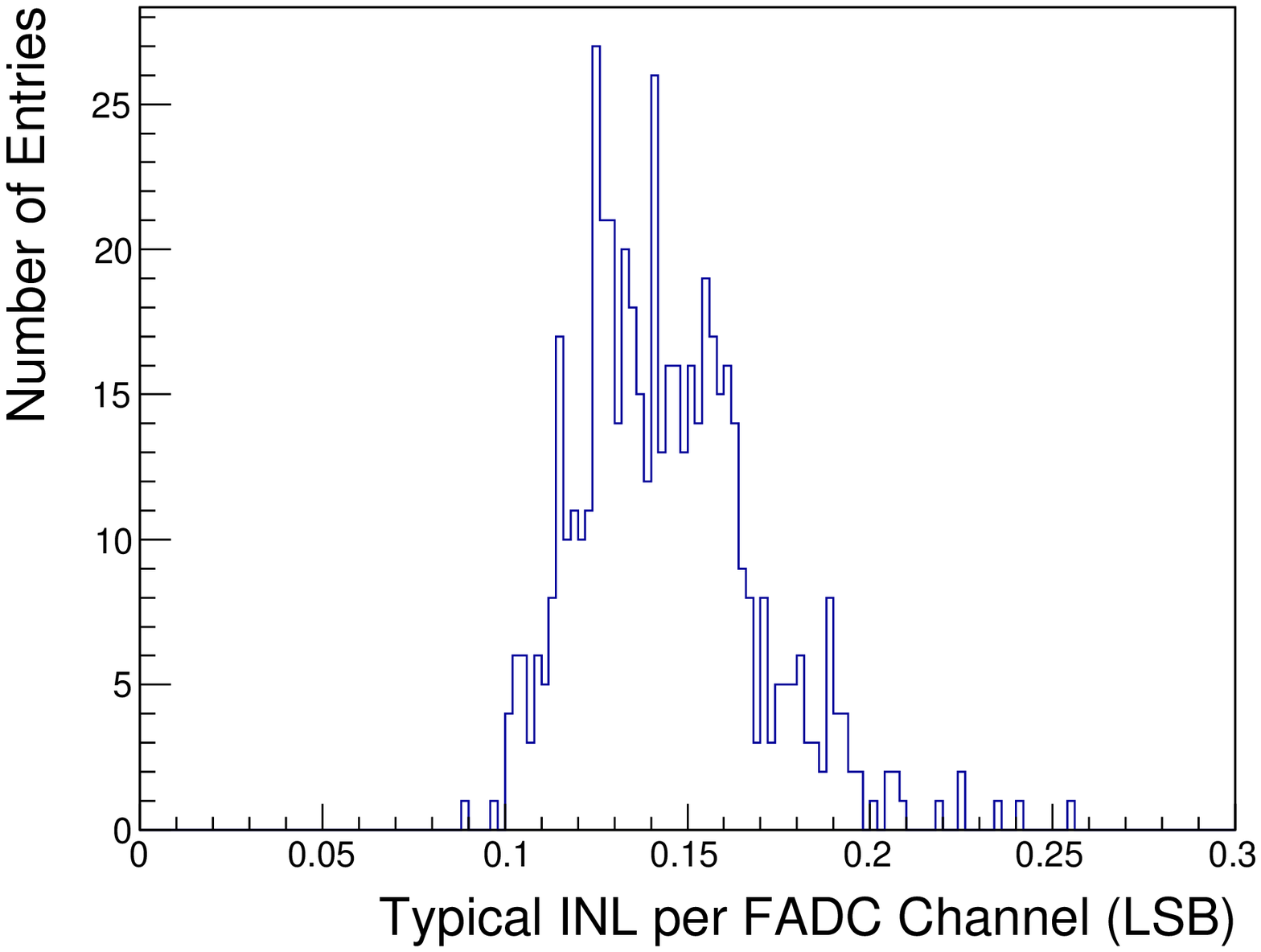}
}

\caption{\label{fig:allcardnonlinearities} \small Non-Linearity measurements of all channels.  On the left, the typical DNL per channel and on the right, the typical INL per channel.}

\end{figure}

Each FADC channel was assessed by calculating the standard deviation of the
obtained DNL and INL values, as shown in Figure \ref{fig:allcardnonlinearities}.  The average of each histogram is used to estimate the typical value of the FADC chip.  The maximum DNL and INL values for each channel were also found to assess the extremes. In our sample of cards, we find that the typical DNL value is $\sim$0.09 LSB, with a maximum of 0.45 LSB.  Similarly for the INL, the typical value is $\sim$0.26 LSB with a maximum of 0.7 LSB.  All values obtained are well within the specifications of the manufacturer, where the typical
maximum DNL value is 0.16 LSB with limits of $\pm$0.6 LSB and typical $INL$ of 0.3 LSB with limits at $\pm$0.9 LSB\cite{FADC_chip}.

%%%%%%%%%%%%%%%%%%%%%%%%%%%%%%%%%%%%%%%%

\subsection{High Frequency Correlated Noise}
\label{noise}
\begin{figure}
\centering{
\includegraphics[width=10 cm]{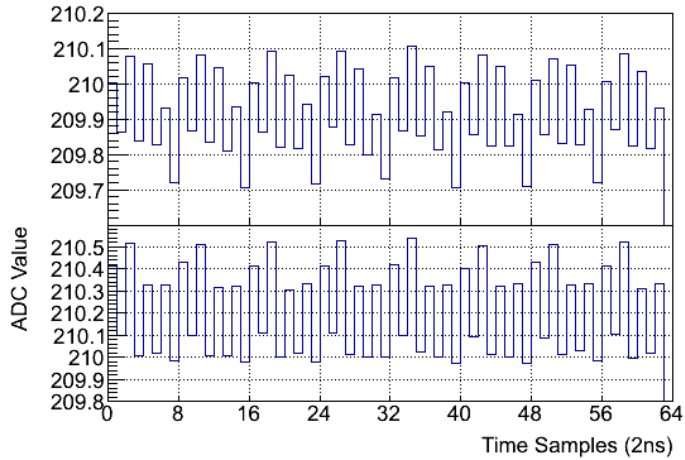}
\caption{Averaged waveforms showing repeated patterns which correspond
         to the frequencies and harmonics of the FADC clocks.} 
\label{fig:clocknoise}
} % caption for the whole figure
\end{figure}

The input noise to the FADC  for the Double Chooz  electronics is low,  with
a measured RMS  of  $\sim$1.2 mV.  Upon averaging of  successively triggered
waveforms,  low amplitude repeating signals  become visible,  as can be seen
in  Figure \ref{fig:clocknoise}.  These  signals  are only observable  when
the waveforms are averaged  (their amplitudes are far below 1 LSB),  so they
occur synchronously to the FADC clocks.  The 16 ns duration of these signals
could correspond  to the oscillation of the main  62.5 MHz clock. The faster
oscillations observed, could be harmonics from the 500 MHz FADC sampling clock.
Correlated noise on the baseline  is not uncommon for boards containing high
speed clocks, and has been observed in many experiments.  The magnitude of
these signals vary from channel to channel and from card to card.

%%%%%%%%%%%%%%%%%%%%%%%%%%%%%%%%%%%%%%%%%%%%%%%%%%%%%%%%%%%%%
\section{Digitisation}
\label{quantisation}

In the digitisation of an analog waveform, two discretisations occur:
\begin{enumerate}
        \item {sampling:} discretisation of the time
        \item {quantisation:} discretisation of the amplitude
\end{enumerate}

If certain conditions are met on both the sampling and quantisation, then the analog signal can be perfectly reconstructed from the digitised waveform. In this case of no-distortion, measurements of parameters such as the integrated pulse charge on the digital waveform yield consistent results as if they had been made with the original analog signal. The condition for sampling was described by Shannon in 1948 and is known as the Sampling Theorem.  Far less well publicised is the condition on the quantisation which was developed by Widrow in the late 1950s \cite{Widrow}.

For the application of Double Chooz, the use of FADCs to record scintillator pulse
shapes from PMTs, the speed of the PMTs and scintillator coupled with the bandwidth
limitation of the Front-End Amplifier and FADC (<200 MHz) ensures that the
condition of the Sampling Theorem is met (sampling at 500 MHz).  The Quantisation
condition, however, requires attention to the amplitude (and form) of the signal
and the analog noise level.

During operation of the \mbox{Double Chooz} far detector, it was observed that each time the DC offset of the FADCs was re-adjusted, not only the pedestal values changed slightly (which was expected), but also the determined gain of each channel appeared to change.  Also an unexpected non-linear energy response was found for each individual channel.  We performed extensive studies, both experimental and by simulation, of the effects in play, and found that the dominant effect was not ADC non-linearity, but quantisation-induced non-linearity.

The charge contained in a signal is calculated by summing the difference between consecutive current samples and the determined baseline.  The
photoelectron signal is extremely variable and the noise induced by the quantisation of the pulse shape is relatively small. The baseline, however, is observed to be extremely stable such that any bias on its knowledge results in a systematic bias on the charge.  The correct determination of the baseline is linked to the analog noise level.  When the noise is large, in comparison to the ADC step (or 1 LSB), the baseline can be well determined. Conversely, when the noise level is low, the baseline position is not well known. Good quantisation of the baseline, we find, is extremely important. The following describes this effect in more detail.

\subsection{Quantisation}
\label{quantisation-math}

 Widrow derived two Quantisation Theorems (QT1 and QT2) linking the signal Probability Density Function and the quantisation step size $q$ which for a Flash ADC would be the voltage (or current) difference between two successive digital values: the ADC step (1 LSB).  The first theorem, QT1, describes the conditions for which there is an unique relation between the statistical descriptions of the input and output signals of the quantiser.  The second, QT2, is a looser condition which, if met, ensures
that at least the moments of the quantised variable are equal to the moments of the sum of the input variable and a uniformly distributed noise. For a good discussion on this subject see \cite{WidrowBook}.

When measuring the signal baseline with a Flash ADC, the signal is the DC
offset plus the analog noise, which, in this case, is observed to be Gaussian
noise, with a standard deviation $\;\sigma_{noise}\;$. If $\;\sigma_{noise}\;$ is larger
than the quantisation step $q$ then QT1 is fulfilled, and complete reconstruction
of the waveform baseline can be made from the digitised version. If
$\;\sigma_{noise}$ > $\frac{q}{2}\;$ then QT2 is fulfilled, and the estimated
mean and variance of the pedestal are equivalent to the input mean and variance.

If QT2 is not fulfilled, the estimated mean and variance of each waveform
baseline will be biassed. In this case, the real DC offset of the baseline will
not be equal to the true offset, and the measured noise level will also be wrong.

The derivations for these biases for several distributions including the Gaussian
case can be found in  \cite{CarbonePetri}.  Here, are reproduced the equations
related to the digital and analog mean and variance of a Gaussian distribution.
The bias observed is related to the analog RMS noise level ($\sigma_n$) expressed
as a fraction of the quantisation step $q$, and the true offset of the waveform
$e_q$. The observed digitised mean offset ($m_e(e_q)$) is:
\begin{equation}
\label{meanbias}
m_e(e_q) = \frac{1}{\pi} e^{-2 \pi^2 \sigma_n^2} sin(2 \pi e_q), \sigma_n \ge 0.3  
\end{equation}

and the digitised output variance is:
\begin{equation}
\label{rmsbias}
\sigma^2 = \frac{1}{12} + \sigma_n^2 - e^{-2 \pi^2 \sigma_n^2} ( (4\sigma_n^2 + \frac{1}{\pi^2})cos(2\pi e_q) -  \frac{1}{\pi^2}  e^{-2 \pi^2 \sigma_n^2} sin^2(2 \pi e_q) ), \sigma_n \ge 0.3  
\end{equation}

From equation \ref{meanbias}, the bias on the measured baseline varies with the
true DC offset of the waveform $e_q$.  The distortion on the waveform
is different for a signal (such as a photoelectron or scintillation
pulse) than for the baseline.  Observable signals use more ADC codes to describe them, and are
inherently variable.  So that we can consider that the main cause of bias on the
measurement of pulse charge occurs due to the waveform baseline (or pedestal). This bias is
simply proportional to the number of samples used to integrate the signal and
the relative value of this bias depends on the signal amplitude. If the signal
amplitude is high, this bias can be small. Conversely if the signal amplitude
is low, then this bias can be significant.

A mismatch between the analog signal, in gain and noise level, and the
quantisation step of a waveform digitiser can lead to a significant non-linearity
on the measurement of pulse charge. A tell-tale sign of this problem is the effect
described in \ref{quantisation}: a shift in the DC offset, which is most
often caused by power cycling of the electronics, results in an apparent shift in the
measured single photoelectron gain.  The first phase of running of the
\mbox{Double Chooz} far detector, suffers from this digitisation problem which is now
experimentaly validated with measurements from a single channel test setup.  

\subsection{Experimental Study of Quantisation Effects}
\begin{figure}
\subfigure{
 \includegraphics[width=7cm]{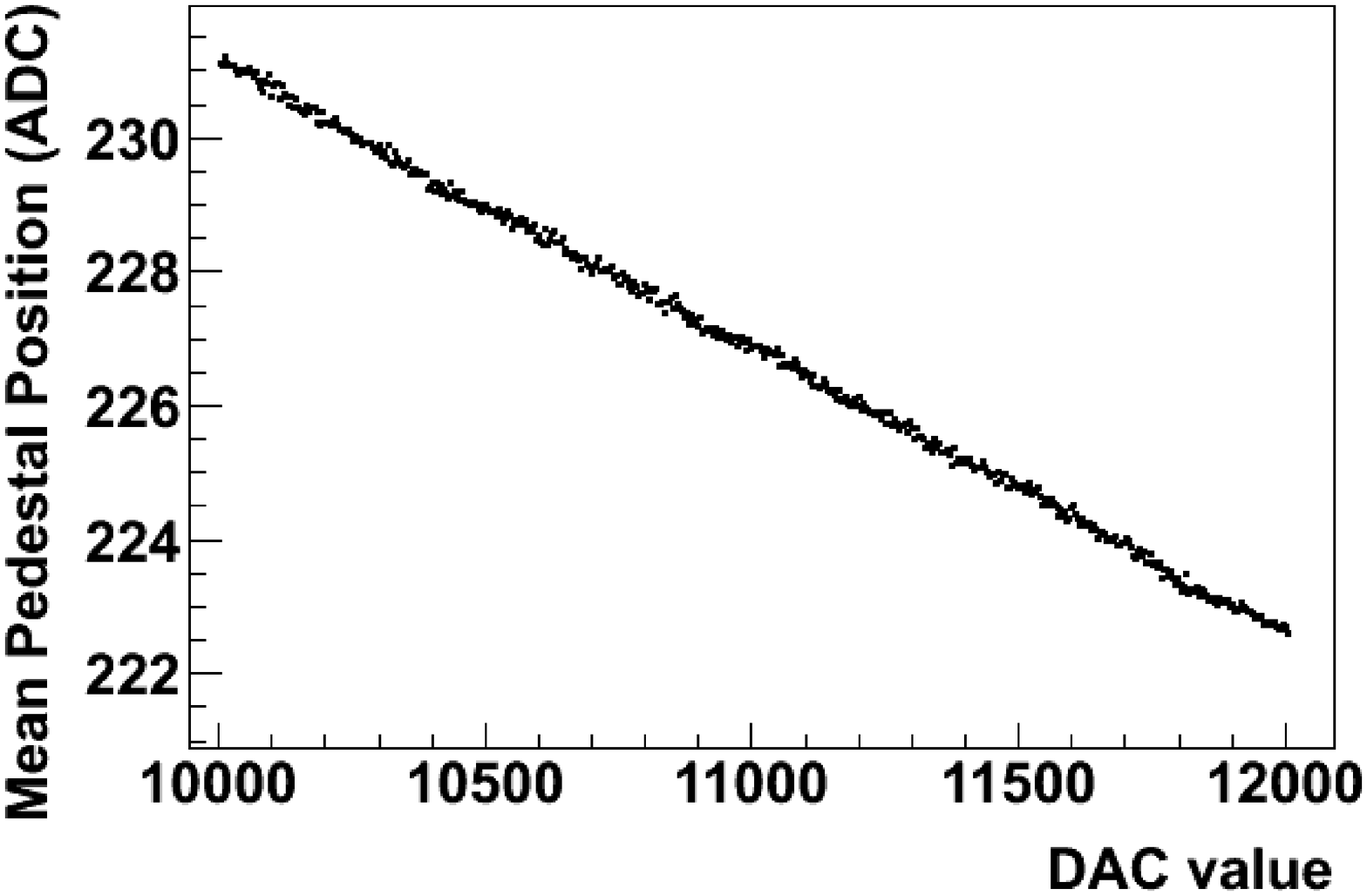} }
\subfigure{
 \includegraphics[width=7cm]{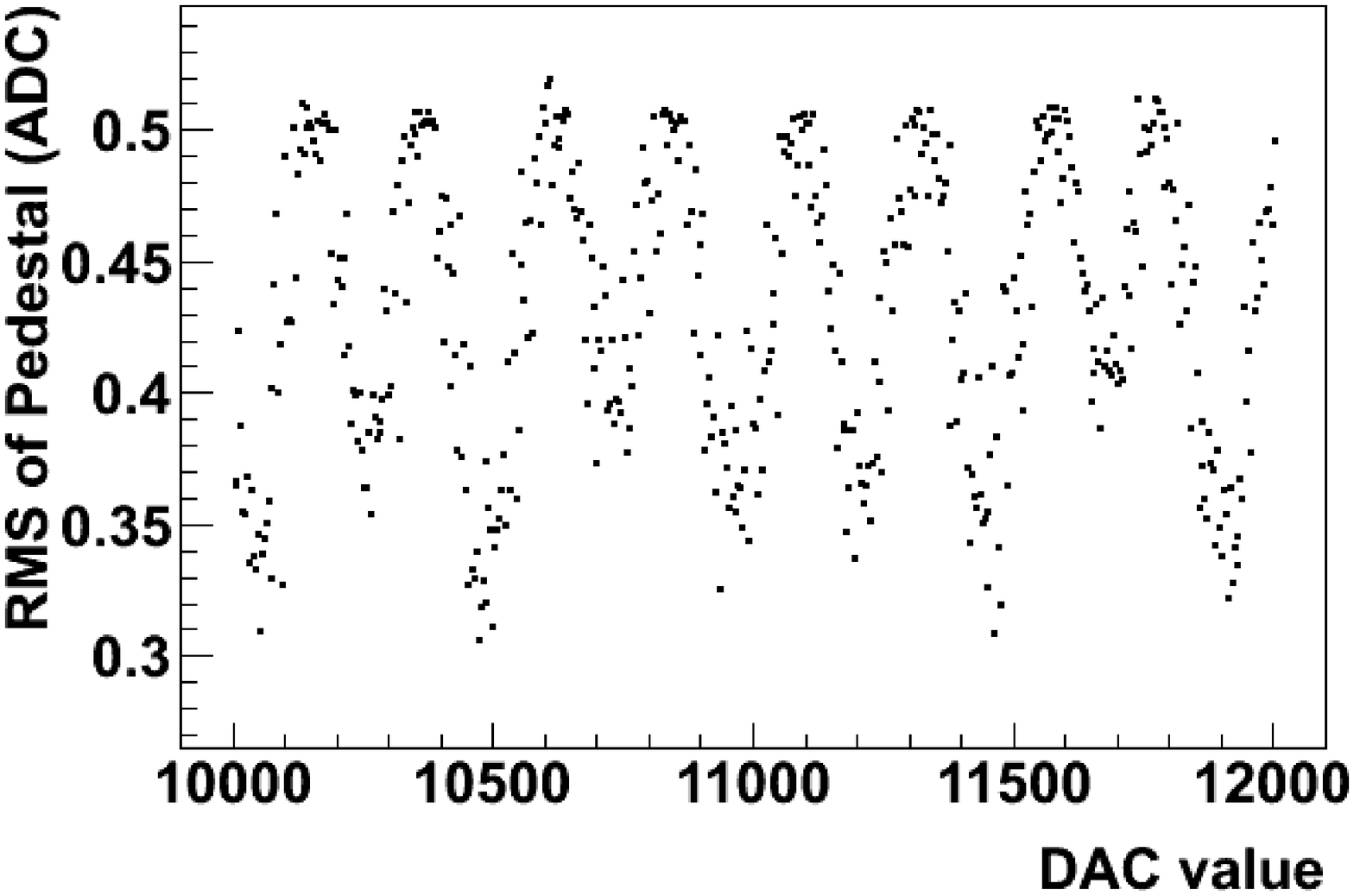} }%
\caption{\label{fig:mean-rms} \small Example of a DAC scan of a Flash ADC
channel crossing 8 consecutive ADC codes. On the left, the estimated
pedestal mean is shown as a function of the input DC offset (DAC) which
is proportional to DC offset voltage and on the right, the waveform RMS. }
\end{figure}

Firstly the DC offset is incremented using the 16-bit controllable DAC whose
value is proportional to DC voltage. For each DAC value the mean and RMS of the
acquired waveforms are plotted in Figure  \ref{fig:mean-rms}.  Oscillatory patterns
are observed for the estimated pedestal mean, as expected from Equation
\ref{meanbias}, and the measured RMS, which oscillates according to Equation
\ref{rmsbias}. 

\begin{figure}
\subfigure{
 \includegraphics[width=7cm]{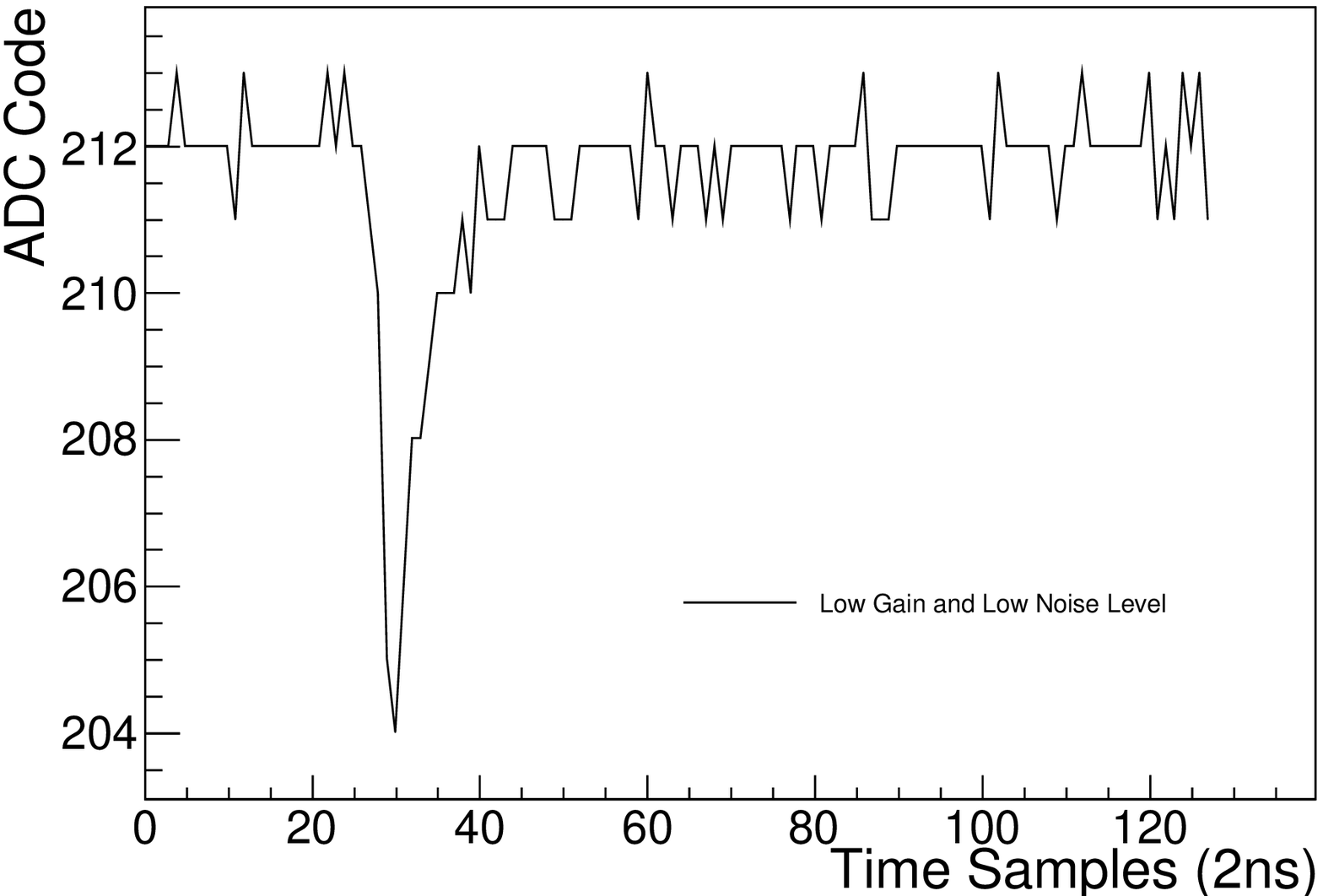} }
\subfigure{
 \includegraphics[width=7cm]{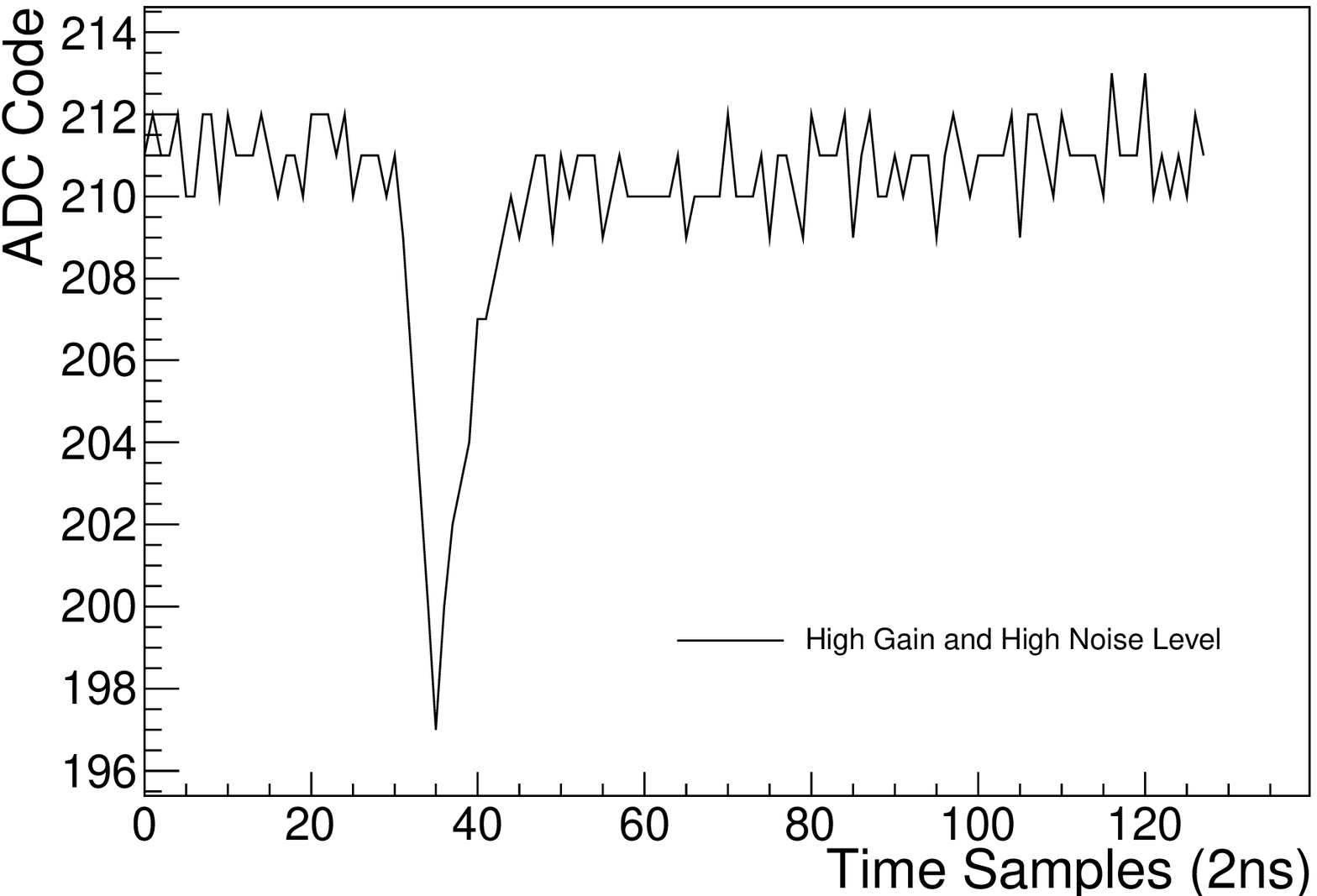} }
\caption{\label{fig:waveforms} \small Examples of waveforms for the case of low gain and low noise (left) and for higher gain and higher noise conditions (right). }

\end{figure}

To test the charge performance, we considered two scenarios; the first is the
standard Double Chooz electronics and gain, with an RMS noise level of 1.2 mV
and a typical mean single photoelectron amplitude of 35 mV, the second is with
a factor of two higher gain and with a RMS noise level of 3 mV. A FADC card is
triggered by the pulse generator each time the LED illuminates. The mean number
of photons per shot is tuned so that all pulses are well-contained within the
8-bit range, and no signals are saturated. For each LED setting, the DC offset
(controlled by the 16-bit DAC) is shifted and 25,000 events are recorded. The
DC offset is moved such that the waveform baseline crosses 3 complete ADC codes.
Data was taken for different levels of LED illumination.  Figure \ref{fig:waveforms} shows two example waveforms of low light-level signals taken under the two conditions of gain and noise.

\begin{figure}
\subfigure{
 \includegraphics[width=7cm]{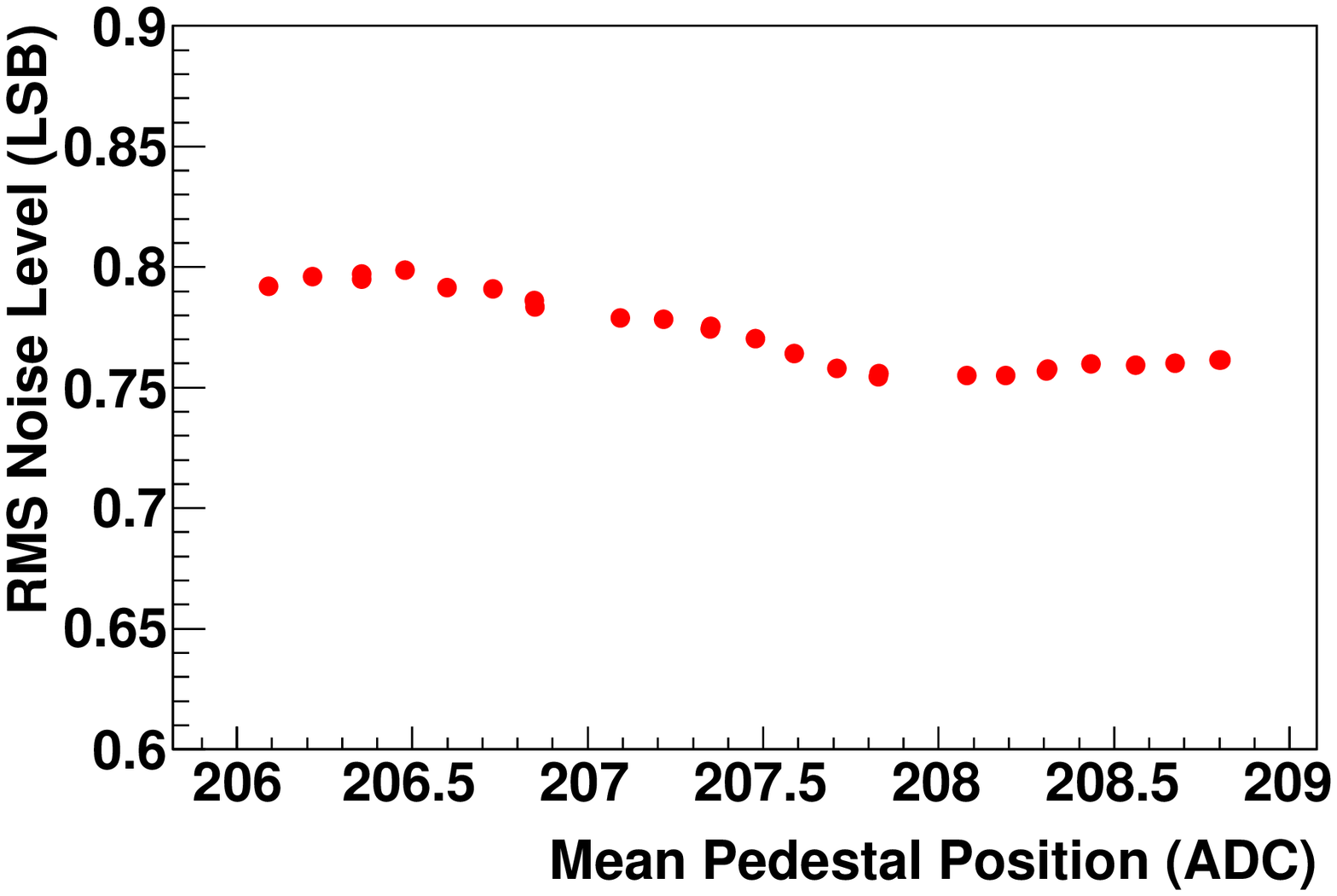}
}
 \subfigure{
 \includegraphics[width=7cm]{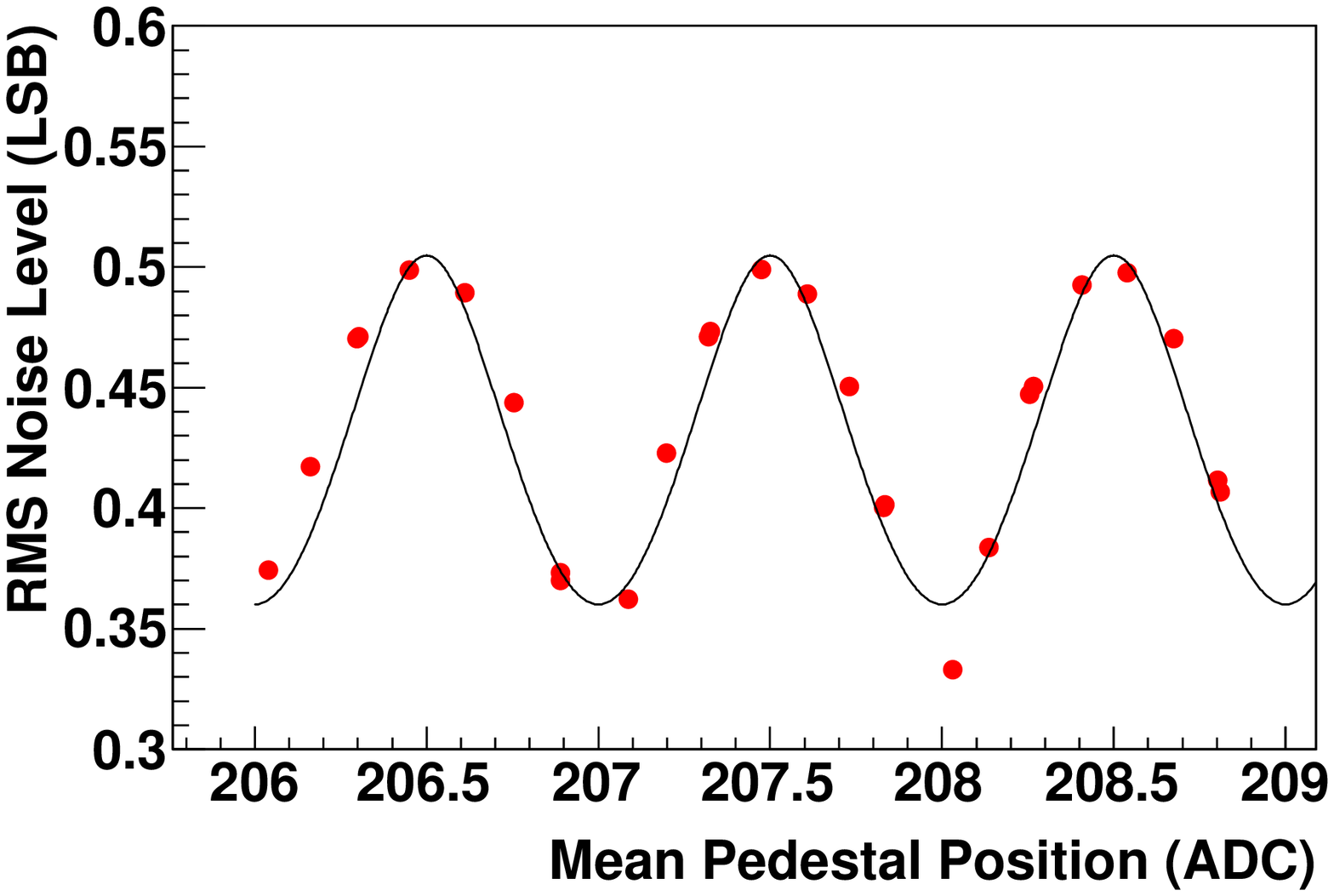}
}
\caption{ \small RMS Noise levels at different DC offsets. Left, the higher analog noise level used for the high gain
runs.
Right, the lower noise case, where a clear oscillatory trend is
observed overlaid with the expected trend assuming a real RMS noise
level of 0.33 LSB.}
\label{fig:lowandhighRMS}
\end{figure}

The RMS noise levels were measured, and are shown in Figure \ref{fig:lowandhighRMS}. At high gain and high noise, a  small variation of the baseline noise level is observed with DC offset, which is most likely due to the variation of code size (DNL).  For the low gain and low noise scenario, a large cyclic varation, corresponding to the transition of the ADC codes, of the RMS noise level is observed. 

\begin{figure}
\subfigure{
 \includegraphics[width=7cm]{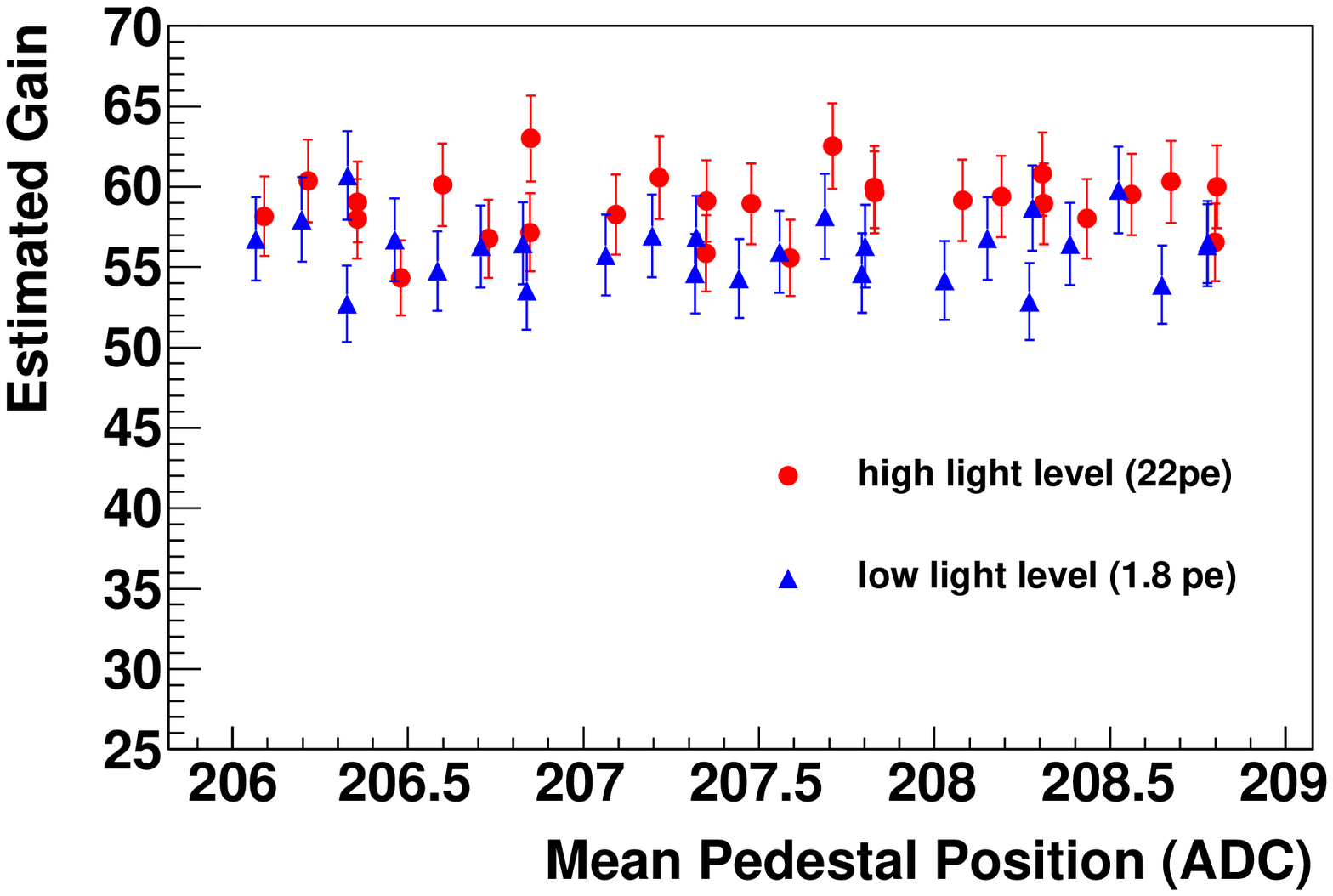}
}
 \subfigure{
 \includegraphics[width=7cm]{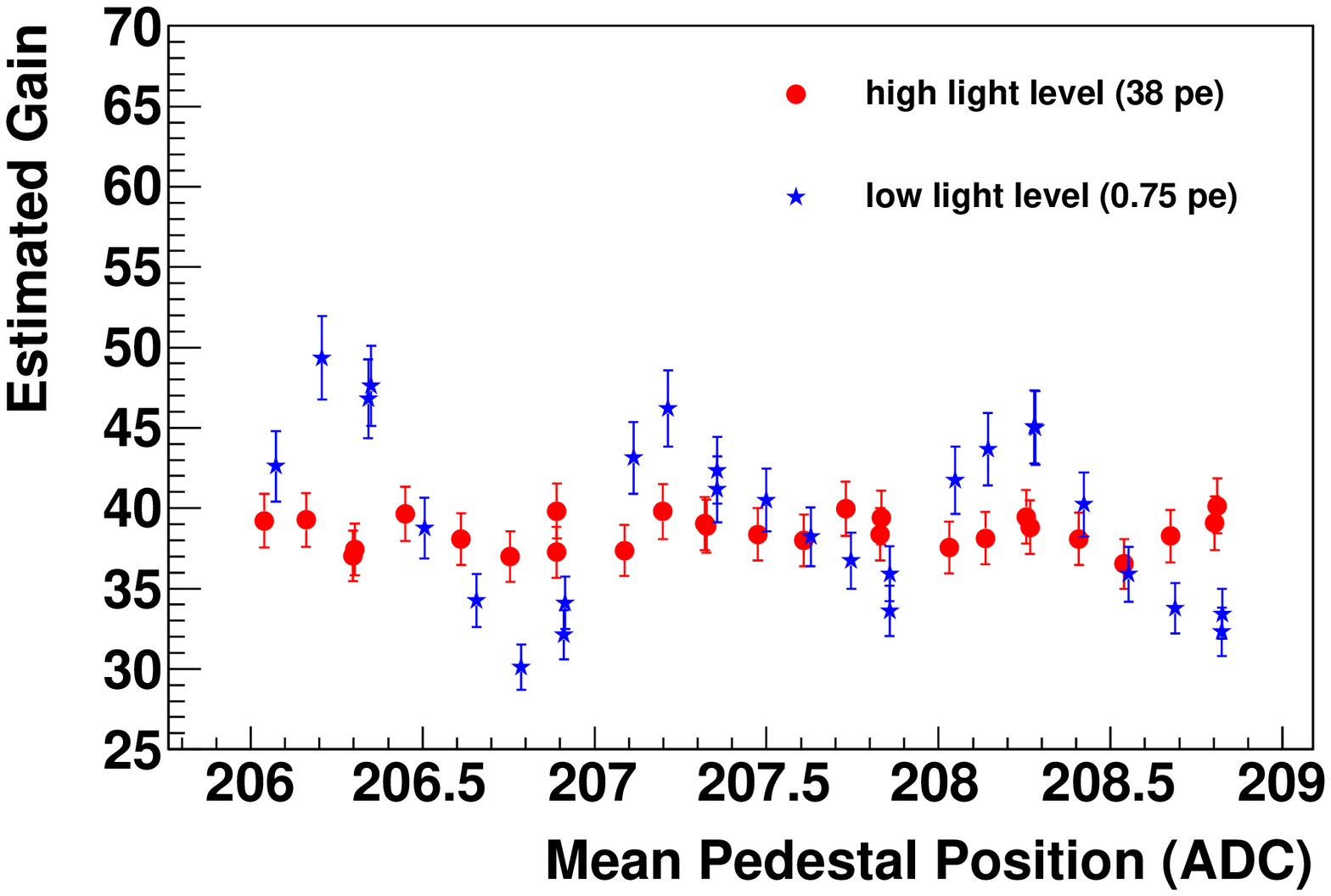}
}
\caption{\small Gain measurements at different DC offsets and light
levels. Left, the high gain and higher analog noise level, resulting
in consistent gain measurements for all light levels (filled circles
and triangles with means of $\sim$22 and 1.8 photoelectrons per LED
shot respectively). Right, the lower gain and noise case, where a clear
oscillatory trend is observed with a greater amplitude for lower light
levels (filled circles and stars with means of $\sim$38 and 0.75
photoelectrons respectively). }
\label{fig:lowandhighgain}
\end{figure}

Figure \ref{fig:lowandhighgain} shows the measured gains for these two conditions. Under good running conditions, the measured gain is expected to be consistent for all values of the DAC offset. The Kuiper test was used, as it is sensitive to cyclic variations, to search for deviations from the expectation that the gain is constant with a significance level of 0.05.  For the high gain and high noise level case, the measured gains are consistent for all values of the DC offset and vary little with light level.  In the low gain and low noise case, however,  large cyclic variations, corresponding to the transitions of ADC code, are observed in the measured gain, at low light levels, becoming less prominent with increasing light levels.  

For the low gain and low noise case, as the signals become larger, the bias on the baseline becomes less significant in comparison to the contained charge, and the oscillations dampen. The shape of the oscillations, the position of minima and maxima, is complicated by the variation of the code size (DNL) and also by the presence of correlated high frequency noise. It is clear that a strong charge non-linearity is present whose magnitude (and sign) is dependent on the pedestal position (DC offset). In this case, the gain measured at the single photoelectron level can be significantly different to that measured at higher light levels.

\begin{figure}

\subfigure{ 
 \includegraphics[width=7cm]{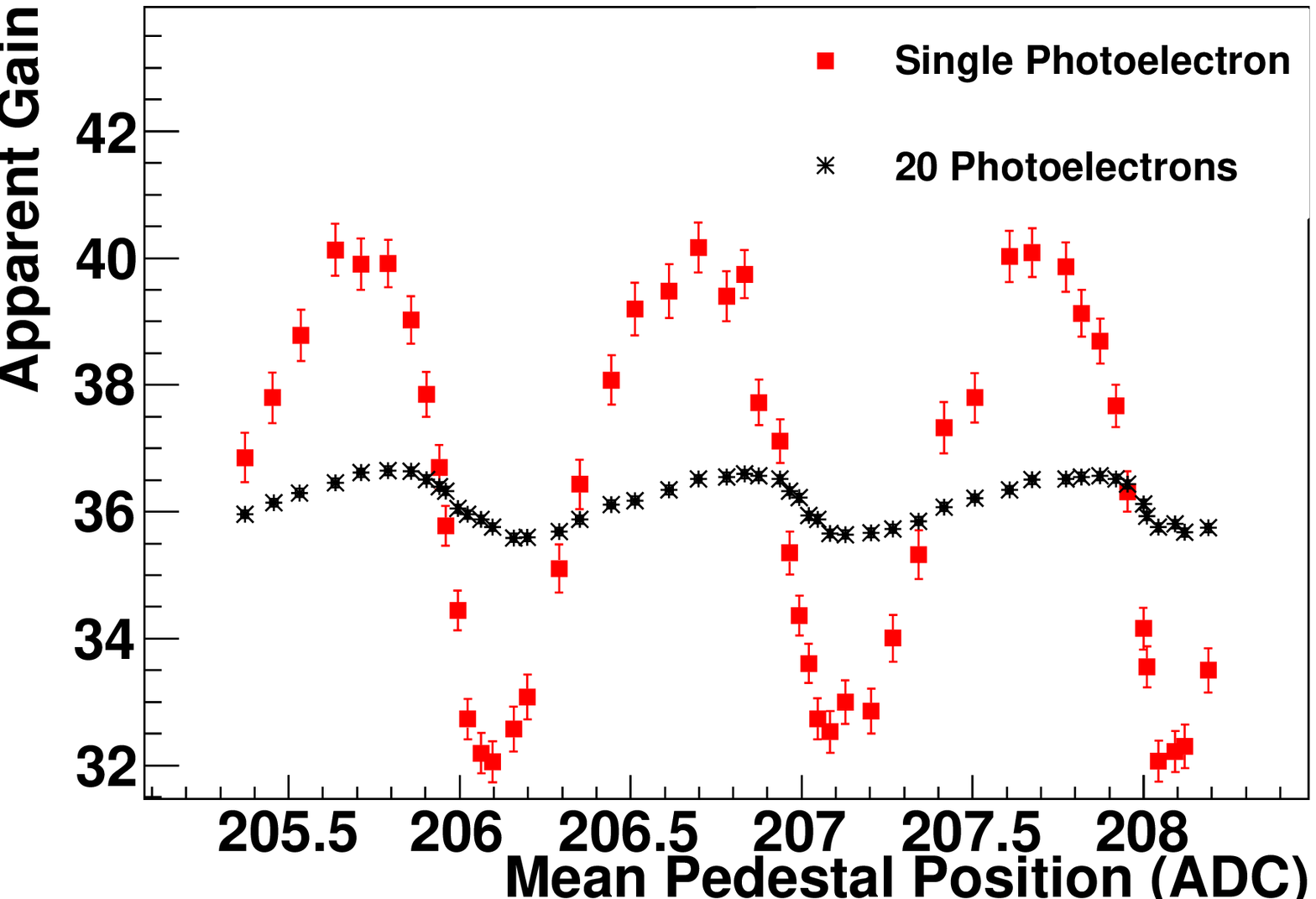}
}
 \subfigure{
 \includegraphics[width=7cm]{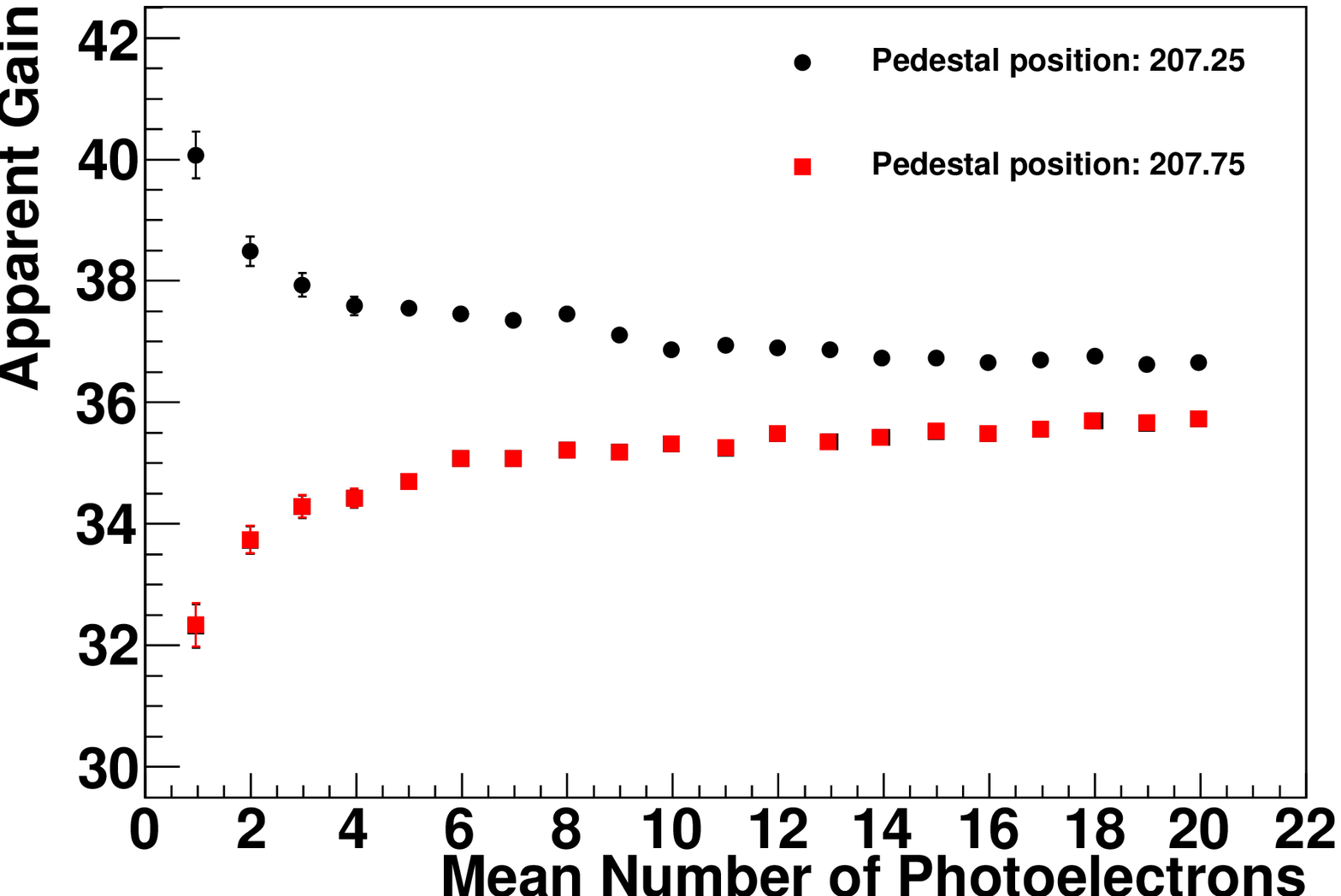}
}
\caption{\small Example from simulation. Left, variation of apparent gain
as a function of pedestal position for 1 photoelectron (filled squares) and
20 photoelectrons (points).   Right, the variation of apparent gain with
increasing number of photoelectrons. Here, shown for two values of pedestal
position, 207.25 (filled circles) and 207.75 (filled squares).}
\label{fig:simfunnel} 
\end{figure}
\clearpage
\section{Simulation}
\label{simulation}

The described effects can be well reproduced by simulation.  Waveforms
are generated, first in an analog sense (in volts) and then 'digitised'
using a voltage-to-ADC transfer function which represents a typical FADC.
The FADC behaviour can be chosen to be perfectly linear or imperfect ie
using measured DNL values or with randomly chosen code widths (whilst
obeying the typical values of maximum INL and DNL of the FADC chip as
measured in Section \ref{dnl}). The analog part of the simulation was
tuned to represent the behaviour of the PMT and Front-End
Amplifier, with typical RMS noise levels. Also included was the
high frequency clock noise, using the observed patterns shown in
Section \ref{noise}.  Single photoelectrons were generated such that
the mean gain as observed by the FADC is $\sim$36 LSB x 2ns. The single
photoelectron charge distribution was assumed to be Gaussian with sigma
of 35\%.  The temporal form of the single photoelectron is given by a
Landau distibution, with parameters tuned to match the actual time
profile of a single photoelectron. The LED pulse profile was generated
assuming a square pulse.

Figure \ref{fig:simfunnel} shows an example of the simulation results.
Here, a perfectly linear FADC was simulated with an analog noise level
of 1 mV and the high frequency noise amplitude of 0.4 mV (approximately
0.1 LSB). A clear oscillation in gain is found for low light levels which
diminishes in strength for higher light levels. Fixing the waveform
pedestal position and plotting the measured gain as a function of
increasing number of photoelectrons shows clearly the charge non-linearity,
shown here for two extreme pedestal positions. Depending on the pedestal
position, the measured gain can either increase or decrease with
increasing light level. Increasing the RMS noise level to 2 mV, as
predicted by QT 2, reduces the charge biases to insignificant levels.

\section{Conclusion}
\label{conclusion}
The waveform digitiser of the Double Chooz experiment was presented. In all
66 cards divided into 5 VME crates are required to form the data acquisition
system of the experiment.  A dedicated firmware allows the synchronous running
of these cards, operating with no deadtime at trigger rates of $\sim$150 events/s.
This firmware also has a flexible readout capability, allowing on-the-fly
decision making on the read-out duration of the waveforms.  These functions
are ideal for rare event searches, such as neutrino reactor experiments,
where the rate of the signals of interest are dominated by the rate of
background events.

The general linearity of the card was assessed through measurements of the
Differential Non-Linearity and Integral Non-Linearity of 67 eight-channel
FADC cards. These measurements aided in a more specific study of the
systematic biases on the determination of the pulse charge from signals
from a PMT tube.

The use of an 8-bit FADC operated in a high dynamic range was described,
where signals vary per channel from 1 to $\sim$50 photoelectrons. Sources of systematic bias were shown for the case where the signal
amplification and analog noise level is low. In this regime, biases from
clock-correlated noise, quantisation and intrinsic non-linearity of the FADC
chip (DNL and INL) are important. These effects were found to be well
reproducable by simulation.  Problems, that can occur if the analog signal and noise
level are not correctly matched to the digitisation step, were explored. We show, the
gain and noise level required to adequately eliminate these problems. 

It is interesting to note that the Sampling Condition (from the Sampling Theorem) is generally taken into account into the design of a waveform digitiser, but the Quantisation condition is not.  It is an error to consider that noise is always a bad thing. Noise is a
dithering agent, and, as such, is necessary. We recommend that
manufacturers provide a means to enable/disable, or, even better, adjust
a source of Gaussian noise to add to the analog input signal, so as to
give the possibility to reach a baseline RMS equal to 0.5 LSB.

%\subsection*{Acknowledgements}

\end{document}